# Differing perceptions on the landing of the rod into the slot

## Chandru Iyer[1] and G. M. Prabhu[2]

[1]Techink Industries, C-42, phase-II, Noida, India, Contact E-mail: chandru_i@yahoo.com
[2]Department of Computer Science, Iowa State University, Ames, IA, USA
Contact E-mail: prabhu@cs.iastate.edu

**Abstract**
In the conventional rod and slot paradox, the rod, if it falls, was expected to fall into the slot due to gravity. Many thought experiments have been conducted where the presence of gravity is eliminated with the rod and slot approaching each other along a line joining their centers, whereby the considerations come strictly under Special Relativity. In these experiments the line of motion is not parallel to either the axis of the rod or the slot. In this paper we consider in detail the two cases when the rod does fall into the slot and when the rod does not fall into the slot, each from the perspective of the co-moving frames of the rod and the slot. We show that whether the rod falls into the slot as determined by Galilean kinematics is also valid under relativistic kinematics; this determination does not depend upon the magnitude of the velocity, but only on the proper lengths and the proper angles of the rod and slot with the line of motion. Our conclusion emphasizes the fact that the passing (or crashing) of the rod as a wholesome event is unaffected by relativistic kinematics. We also provide a simple formula to determine whether or not the rod passes through the slot.

## 1. Introduction

Einstein [1] showed that the Lorentz transformation equations describe the relation between space-time event coordinates (x, y, z, t) and (x', y', z', t') of the same event from two inertial frames under the principles of equivalence of inertial frames and constancy of the speed of light. The Lorentz transformation equations, according to the principle of relativity, describe differences in observation depending on motional speed. If the length of an object at rest is $L_0$, then when it moves relative to a coordinate system with a speed $v$ it is expected to appear contracted from $L_0$ to L according to the relationship:

$$L = L_0 / \gamma$$

where $\gamma = \dfrac{1}{\sqrt{1 - v^2/c^2}}$ and $c$ is the speed of light in vacuum. In a co-moving coordinate system, the length L′ will still appear to be equal to $L_0$. Over the years, the Lorentz transformation equations have given rise to a number of paradoxes. The simplest one-dimensional version is the Pole in the Barn problem [2] which focuses entirely on the question of simultanety. Later versions include the rod and hole paradox, first published in [3] and later in [4].

While Rindler [3] advanced the idea of differing perceptions of rigidity to explain the observations by the rod and the slot, Shaw [4] felt that as the rod starts to 'fall,' it experiences accelerated motion. Under those conditions it will look curved from one frame and straight from another. Shaw [4] further opined that the paradox can be taken out of the purview of gravity and accelerated motion by considering a different problem where there is no gravity and therefore more strictly under the special theory of relativity. He proposed a slot and a rod approaching each other (as seen from a base inertial frame F). The rod has a large velocity along the x-axis with respect to F, while the slot has a small velocity in the z-axis (with respect to F) moving in such a way that their centers coincide at an instant t = 0 in frame F. He showed that while the slot perceives that the rod fell into it with their lengths fully aligned, the rod perceives that it went into the slot in an inclined way and thus could go into the slot even though it (the rod) was longer than the slot.

Grøn and Johannesen [5] simulated the falling of the rod into the slot by developing a computer program which numerically transforms the coordinates from one frame to another by the Lorentz transformation equations and displays the view from the co-moving frames of the rod and the slot by computer graphics. The graphics vividly show that the shape and inclination of the rod is very different as observed from the two inertial frames. This confirms Rindler's conjecture that the perceptions of 'rigidity of the rod' may vary as seen by the two frames. The question of whether the nature of a physical effect depends upon the frame of reference from which the said effect is observed is addressed in [5] by emphasizing the need for a relativistic theory of elasticity. Grøn and Johannesen note that while a breakage in the rod is a physical effect, the bending of the rod is not a physical effect, as different reference frames perceive the extent of this bending differently.

Marx [6] observed that as the velocity of the rod increased, the observation from the co-moving frame of the slot would be rod rotated and contracted but the line of motion of the end points of the rod would remain the same. This indicated that the rod's passing or not passing through the slot was not affected by the magnitude of the approach velocity. Martins [7] emphasized the point that any one-to-one space-time coordinate transformation, even if other than Lorentz, does not lead to a contradiction; a point-to-point collision between objects as observed in one frame necessarily transforms to a collision in the other frame also, even if the space-time coordinates are different in the two frames for that collision. Thus the passing of the rod or otherwise is consistent in both the frames even if the temporal sequences of certain events are relative.

This paradox was recently re-visited in [8]. According to [8] the car and hole paradox described in [9] and the rod and hole paradox described in [3 and 4], do not give rise to any contradiction when explained on the basis of stress propagation – that is, the difference in observed speed of stress propagation compensates exactly for the differences in observed length.

In this paper we present a variant of the rod and slot paradox with motion in two directions but only with constant velocity. The pedagogical contribution of our paper shows how the non-invariance of angles and proper lengths comes into play. In this scenario the line of motion (that is, the line joining the centers of the rod and slot) is not aligned with either the x-axis or y-axis, there is no gravity, and thus no stress or propagation of stress. We show that whether the rod passes through the slot (or not) is determined only by four quantities: (i) proper length of rod, (ii) proper length of slot, (iii) proper angle between axis of rod and line of motion, and (iv) proper angle between axis of slot and line of motion. We provide a simple formula to determine whether the rod passes through the slot. This determination is independent of the approach velocity and relativistic kinematics and remains the same under Galilean kinematics.

## 2. The rod and the slot approaching under no gravity

Imagine a sheet metal located somewhere in space where there is no gravity. Let the sheet metal be on the x-y plane and contain a slot of length *l*. Further consider a rod also of length *l* traveling in such a way that it is trying to pass through the slot from one side of the x-y plane to the other side. For this purpose it has a velocity component along the length of the rod, say the x-axis, and another along the negative z-axis. This is similar to a plane landing on a runway. The predominant velocity is on the x-axis and we have a very small velocity on the negative z-axis. Under our assumption of no gravity we also eliminate along with it the effects of stress, stiffness, and propagation of stress. Taking the center of the slot as the origin of the system, we make the following initial observations (as seen from the rod's perspective and shown in Figure 1):

- The initial condition is such that the center of the rod is located at $x = -a$, $z = +b$.
- The y coordinate is not relevant to the problem as all events occur on the x-z plane along the axis of the slot.
- There is no gravity or any external force.
- The velocity $v_x$ of the approaching slot along the x-axis is large and comparable to $c$.
- The velocity $v_z$ of the approaching slot in the z direction is small.
- In order to land perfectly into the slot we assume the ratio $a/v_x$ is equal to $b/v_z$.
- This will make the rod land exactly into the slot after a time $t = a/v_x = b/v_z$.
- It is also possible to make both 'b' and '$v_z$' tend to zero while maintaining $a/v_x = b/v_z$.

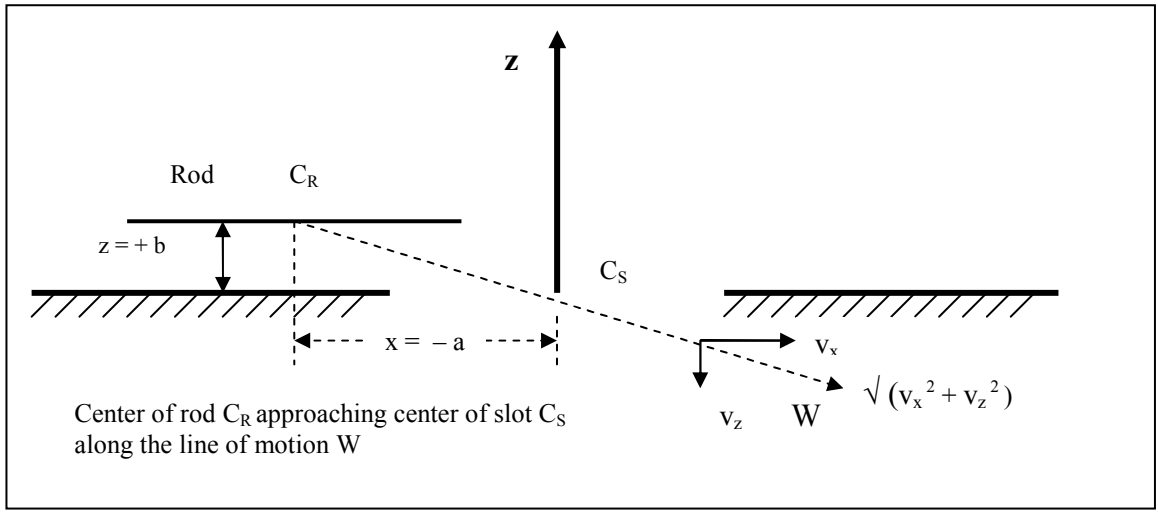

**Figure 1. The rods' perception: Initial conditions of the rod and the slot**

## 3. The perceptions of the rod and the slot

The axis of neither the rod nor the slot is collinear with the line of the relative velocity (W). We use the following notation (shown in Figures 2 to 5) to designate the relevant angles.

Φ: Acute angle between the axis of the rod and W (the line of motion), as perceived from the co-moving frame (R) of the rod.
θ : Acute angle between the axis of the rod and W as perceived from the co-moving frame (S) of the slot.

α : Acute angle between the axis of the slot and W as perceived from the co-moving frame (S) of the slot.
β: Acute angle between the axis of the slot and W as perceived from the co-moving frame (R) of the rod.

φ and α are the 'proper' angles in the respective co-moving frames. From the Lorentz contraction it is well established that from the other frame these two angles will look larger (unless they are zero or 90 degrees, in which case the two angles will appear the same from the other frame). Thus

$\theta > \Phi$  and  $\beta > \alpha$        ------------------------------------------------------------ (1)

The relation between the angles is

$\tan(\theta) = \gamma \tan(\Phi)$
$\tan(\beta) = \gamma \tan(\alpha)$

where $\gamma = \dfrac{1}{\sqrt{1 - (v_x^2 + v_z^2)/c^2}}$

In the following discussions we consider three cases.

**Case I:** $\theta = \alpha$ (The slot sees the rod landing with both their axes aligned)

In this case (Figure 2) the rod goes into the slot with their axes fully aligned and the rod is smaller than the slot (all perceptions are from frame S).

The rod perceives this same case in a different way. For this we consider the relationship between $\Phi$ and $\beta$ when $\theta = \alpha$. When we incorporate the inequalities of (1) above with this equality we get $\Phi < \theta = \alpha < \beta$; thus we have $\Phi < \beta$ for the case when $\theta = \alpha$.

In other words, when the slot sees an aligned landing, the rod sees a landing with the leading edge of the rod tilted towards the slot (Figure 3); we call this a favorable alignment because this facilitates the rod passing through the slot even though the rod is 'longer' than the slot.

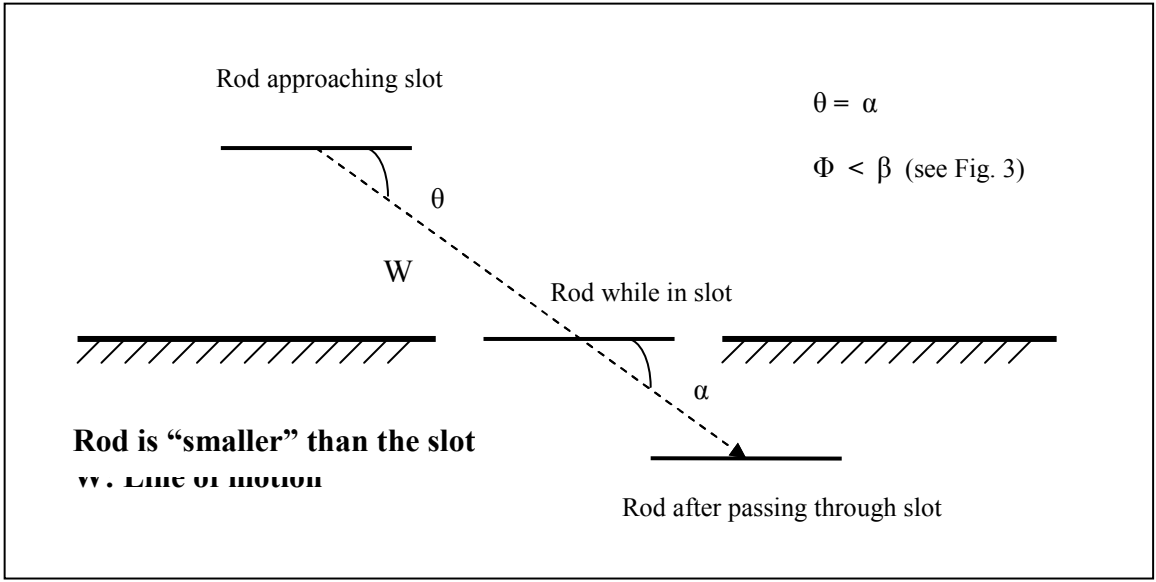

**Figure 2. The slot's perception: Rod went through the slot**

Figures 2 and 3 depict the interpretations of the rod and slot in Case I when the rod did pass through the slot.

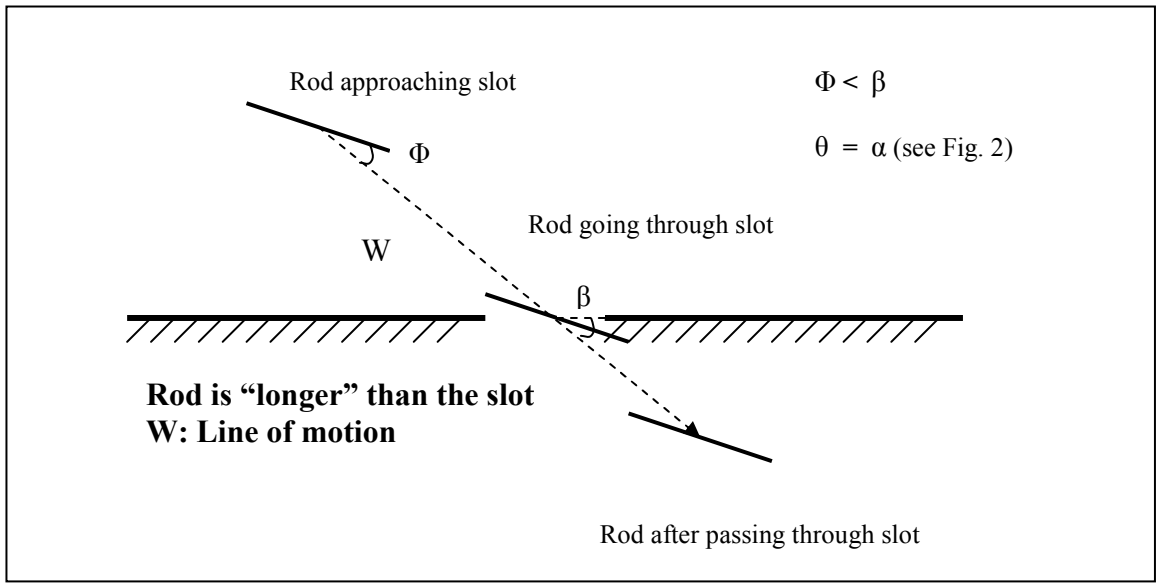

**Figure 3. The rod's perception: Rod went through the slot**

**Case II:** $\Phi = \beta$ (The rod sees itself aligned with the slot – Figure 5)

In this case the slot will see (Figure 4) that the rod is approaching with its leading edge tilted upwards, $\theta > \alpha$. The rod does not go into the slot (Figure 5) because according to the rod it was longer than the slot; but according to the slot (Figure 4), the rod, even though smaller, was unfavorably aligned.

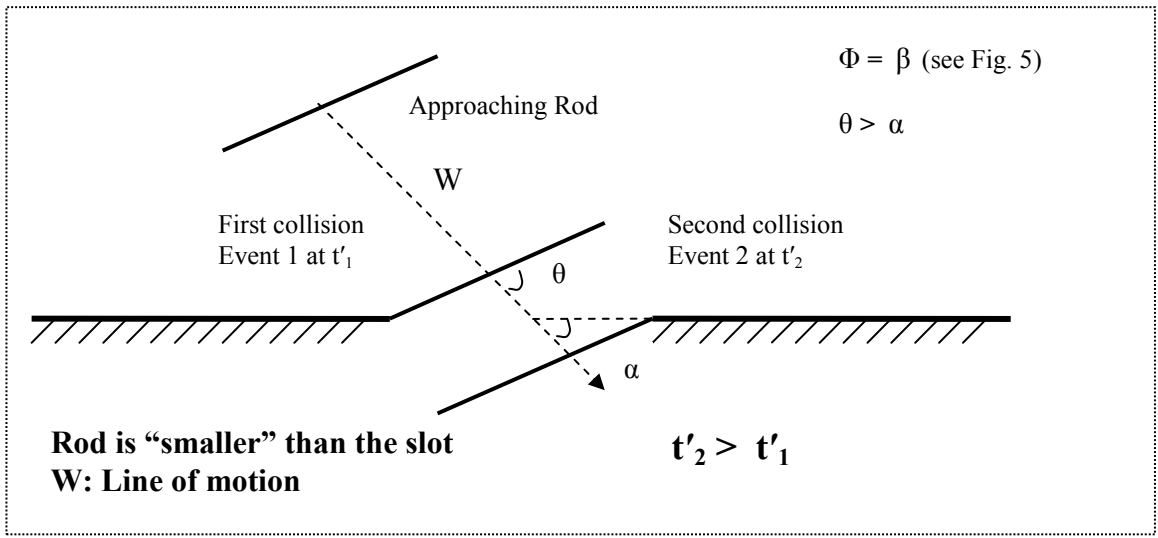

**Figure 4. The slot's perception: Rod did not go through the slot**

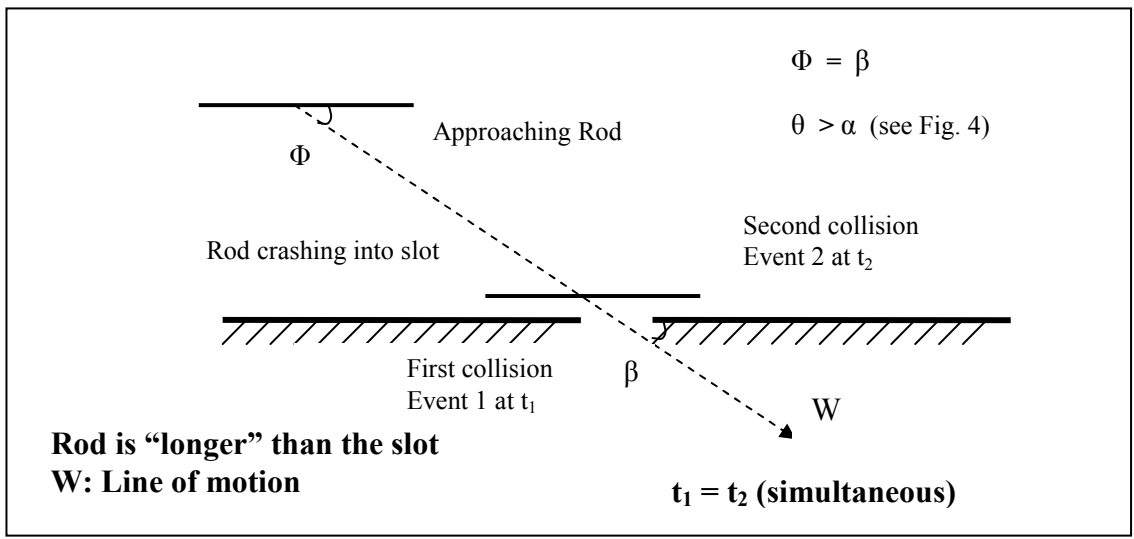

**Figure 5. The rod's perception: Rod did not go through the slot. It crashed.**

Figures 4 and 5 depict the interpretations of the rod and slot in Case II when the rod did not go through the slot, that is it crashed.

Thus the rod and the slot disagree on their alignments as well as their lengths in both the cases. The rod passing through the slot is a reality in Case I; the rod **not** passing through the slot is a reality in Case II. On these realities the rod and slot agree in both cases. Only the reasons for the occurrences are either assigned to inequality in lengths or favorable/unfavorable alignments.

But in Case II, the collision of both edges of the rod onto the slot is simultaneous according to the rod (Figure 5). These two events are not simultaneous according to the slot (Figure 4). This is somewhat intriguing as conventional thinking precludes the second collision after the first collision. However, the stress or disturbances from the trailing edge collision cannot 'travel' to the leading edge, before the 'second' collision occurs. This aspect has been dealt with in detail in a recent article [10].

**Case III:** $\Phi = \alpha$

Now we consider a situation where the proper angles between the axis of the rod and the line of motion and that of the slot with the line of motion are equal. This situation is not a perception from any of the inertial frames, but it has significance as this case maintains the symmetry between the two linear objects and the two corresponding frames. In this case, when we assume the proper lengths of the rod and the slot to be the same, we observe that the rod sees itself as entering the slot in a favorable alignment. But it also sees that its length (i.e., the rod's) is longer than the perceived length of the slot. The favorable alignment and the higher length compensate for each other and the rod goes through the slot by just a whisker. This result is the same under Galilean considerations when the proper angles are equal and the proper lengths are equal.

Similarly, the slot perceives the rod to be smaller and unfavorably aligned; both these aspects offset each other and the rod goes through the slot by just a whisker. These results are as expected because the condition $\Phi = \alpha$ maintains the symmetry between the two linear objects.

**4. Galilean kinematics**

When we consider the problem from the Galilean model, with no distortion in the lengths or the angles, we find that the projection of the slot parallel to the line of motion plays no role in the passing of the rod through the slot. Similarly, the projection of the rod parallel to the line of motion does not contribute to a minimum slot length requirement. Thus we find whether the rod passes through the slot or not is determined only by the projections of the rod and slot on the line perpendicular to the line of motion; that is when the rod is trying to pass through the slot, it is as if the projection of the rod perpendicular to the line of motion is trying to pass through the projection of the slot perpendicular to the line of motion. In other words when $L_R \sin \Phi < L_S \sin \alpha$ the rod falls into the slot. This is illustrated in Figure 6.

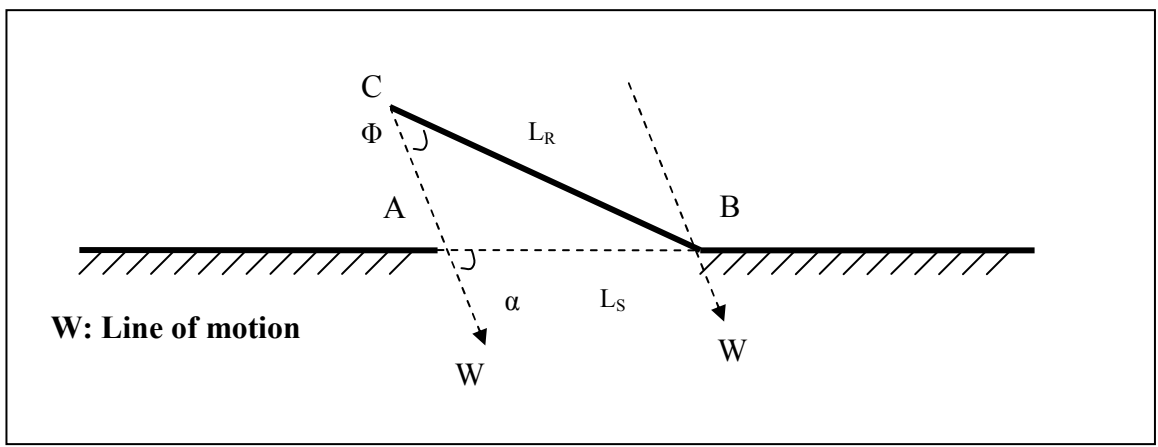

**Figure 6. Galilean kinematics: The rod just passing through the slot**

Here the rod lands at an angle and when the leading edge of the rod coincides with the front edge of the slot, the rod and slot form two sides of a triangle. Further we assume that the rod just passes through the slot; this means the line joining the trailing edge of the rod and the back edge of the slot coincides with the line of motion and forms the third side of the triangle. When we apply the law of sines for triangle ABC, and observe that $\sin \alpha = \sin(180 - \alpha)$, we obtain the equation $L_R \sin \Phi = L_S \sin \alpha$. This is for the situation when the rod just manages to pass through. Thus we derive the condition that whenever $L_R \sin \Phi < L_S \sin \alpha$, the rod falls into the slot.

When the line of motion coincides with the axis of the rod, a pinhole slot of almost zero length is sufficient to let a very long rod pass through; in this case $\Phi = 0$ and $L_R \sin \Phi = 0$. In our presentation, we have assumed that the centers of the rod and slot approach each other along the line of motion. In the conventional rod and slot paradox, the center of the rod is just a little bit above the center of the slot and gravity is required to aid the fall.

### 5. Relativistic kinematics

We would expect that under relativistic kinematics, from the perception of any one frame the same considerations as above to be valid with the proper lengths and proper angles replaced with the lengths and angles perceived by that frame. From the slot's co-moving frame $L_S \sin \alpha$ is unaltered, as this frame will perceive the proper length of the slot, and the angle $\alpha$ without any distortion. The length of the rod $L_R$ and the angle $\Phi$ will be perceived as different by the slot to a contracted length $L_{RA}$ and angle $\beta$ respectively, but in such a way that $L_R \sin \Phi = L_{RA} \sin \beta$.

Thus we find the condition $L_{RA} \sin \beta < L_S \sin \alpha$ is identical to the condition $L_R \sin \Phi < L_S \sin \alpha$. A similar logic can be presented for the co-moving frame of the rod. In other words the condition that determines whether the rod will pass through the slot is unchanged as we switch from Galilean to relativistic kinematics in either of the inertial frames.

### 6. Summary

The above three cases are interesting in that apart from their proper lengths, the orientation of the rod and slot with the line of motion plays a key role in determining whether the rod falls into the slot. We have shown whether the rod passes through the slot is easily established by evaluating if $L_R \sin \Phi < L_S \sin \alpha$. This determination is unaffected by relativistic kinematics and the magnitude of the velocity. The four (proper) quantities $L_R$, $L_S$, $\Phi$, and $\alpha$ conclusively determine the result both under Galilean and relativistic kinematics.

Thus we see that when we have a planar motion, there is distortion of angles as well as linear dimensions and the reasons attributed for physical occurrences are not only assigned to dimensions but also to angles. The physical occurrence itself is observed to be the same from different inertial frames; however, they assign different reasons to the observations.

**Acknowledgments:** The authors would like to thank the reviewers for greatly improving the presentation of the paper.